\documentclass[aps,prl,twocolumn,amsmath,amssymb,nofootinbib,floatfix]{revtex4-2}

\usepackage{graphicx}   
\usepackage{bm}         
\usepackage{hyperref}   
\usepackage{xcolor}     

\begin{document}

\title{Collective Dynamics and Topological Locking in Knotted Ring Polymers:\\
A Novel Phenomenological Theory}

\author{Anna Lappala}

\affiliation{Department of Molecular Biology and Mass General Hospital, Harvard University, Boston, 02114, USA}

\date{\today}

\begin{abstract}
\textbf{We present a novel phenomenological theory describing how topological constraints in prime-knot ring polymers induce collective (cooperative) modes of motion}. In low-complexity knots, chain segments can move quasi-independently. However, as the crossing number increases, the ring's degrees of freedom become \emph{collectively} coupled: distinct arc segments must move in coordinated, out-of-phase patterns to preserve the knot. We formulate this using an arc-based model in which each crossing imposes constraints that generate a coupling matrix among subchain displacements. We show how strong couplings emerge at higher knot complexity, eventually leading to topologically driven \emph{dynamical arrest}.  We demonstrate that \textbf{torus and twist knots belong to distinct universality classes} of topologically driven dynamical arrest: torus knots exhibit a gradual, stretched-exponential slowdown, while twist knots undergo a sharp, jamming-like transition. These findings establish a topological control parameter for relaxation dynamics, independent of steric effects or bending rigidity. Our results offer a unified framework connecting number of conformations, cooperative motion, and final arrested states, thus extending our fundamental understanding of entanglement in soft matter systems.
\end{abstract}

\maketitle


\textbf{Introduction.}---Knotted ring polymers exhibit striking physical behaviors due to topological constraints that restrict how segments can move without passing through each other \cite{Grosberg2007, Matthews2010, Dai2015, Orlandini2016, VargasLara2017, Tubiana2011, Tubiana2018, Kholodenko1992, Kholodenko1996, Thompson2025}. Recent simulation and experimental studies reveal that \emph{complex} knots (with larger average crossing numbers) can undergo partial or complete \emph{dynamical arrest} \cite{Bao2003,Zhang2019}. Yet the microscopic mechanism underlying cooperative motions---where one segment ``pushes'' and another ``pulls''---remains elusive. 

Here, we develop a \emph{new phenomenological theory} that explicitly encodes the coupling among \emph{arcs} (subchains in the knot diagram) induced by topological constraints. We show that, at low complexity, these couplings are negligible and motion appears relatively uncoordinated. At higher crossing numbers, stronger constraints force segments into large-scale \emph{collective} displacements, ultimately leading to topologically locked states with few accessible conformations. Our arc-based framework thus unifies (i)~the emergence of correlated principal-component modes, (ii)~the collapse in number of metastable conformations, and (iii)~dynamical arrest at large crossing number.


\textbf{Arc-Based Model.}---Consider a prime knot with $n_c$ minimal crossings.  In its minimal diagram, there are $2 n_c$ \emph{arcs}, each of which we treat as a subchain of length $L_i$ (in monomer units).  Let $\mathbf{R}_i$ represent a collective coordinate for arc $i$ (e.g.\ a center-of-mass or principal displacement), and let $\boldsymbol{\phi}_i$ encode its internal shape degrees of freedom.  The ring's overall configuration is then 
\begin{equation}
\{\mathbf{R}_1,\ldots, \mathbf{R}_{2n_c};\;\boldsymbol{\phi}_1,\ldots, \boldsymbol{\phi}_{2n_c}\}.
\end{equation}

\emph{Topological constraints} appear through crossing points in the diagram.  In three dimensions, these arcs cannot ``pass through'' each other without violating the knot type.  We capture this by introducing a \emph{constraint energy}:
\begin{equation}
F_\mathrm{topology} \;=\; 
\sum_{\langle i,j\rangle} \Delta(\mathbf{R}_i,\mathbf{R}_j),
\label{eq:constraint}
\end{equation}
where the sum runs over pairs of arcs $(i,j)$ that share a crossing.  $\Delta(\mathbf{R}_i,\mathbf{R}_j)\rightarrow \infty$ if arcs $i$ and $j$ come too close to passing strands, thereby forbidding such topological transgressions.

We further include bending costs and entropic contributions:
\begin{equation}
F_{\mathrm{bend}} + F_{\mathrm{ent}} \;=\; 
\sum_{i=1}^{2n_c} 
\Bigl[ \kappa \,\mathcal{B}(L_i,\boldsymbol{\phi}_i)
- T\, S_{\mathrm{conf}}(L_i,\boldsymbol{\phi}_i) \Bigr].
\label{eq:free}
\end{equation}
Here $\kappa$ is the persistence length and $S_{\mathrm{conf}}$ represents the internal configurational entropy of arc $i$. $\mathcal{B}(L_i,\boldsymbol{\phi}_i)$ represents the bending energy contribution, which depends on the length $L_i$ of the arc and its internal shape degrees of freedom, $\boldsymbol{\phi}_i$. It captures the elastic deformation cost of maintaining the shape of each arc under the constraints imposed by the knotted topology. Altogether, Eqs.\ (\ref{eq:constraint})--(\ref{eq:free}) define an effective \emph{free energy} $F$ for the knotted ring.


\textbf{Coupling and Cooperative Modes.}---Linearizing around a reference conformation $\mathbf{R}_i^{(0)}$, we write 
\begin{align}
F_{\mathrm{coll}} \;\approx\; 
\frac12 \sum_{i,j=1}^{2n_c}
\bigl(\mathbf{R}_i - \mathbf{R}_i^{(0)}\bigr)^T \,K_{ij}\,
\bigl(\mathbf{R}_j - \mathbf{R}_j^{(0)}\bigr),
\label{eq:F_coll}
\end{align}
where the \emph{coupling matrix} $K_{ij}$ encodes how displacements of arc $i$ induce compensatory adjustments in arc $j$.  In a \emph{tight} or \emph{complex} knot (large $n_c$), $K_{ij}$ has strong off-diagonal terms (due to multiple crossing constraints).  Conversely, at low $n_c$, $K$ is sparse, permitting relatively uncoordinated motion.

From Eq.~\eqref{eq:F_coll}, the \emph{normal modes} satisfy
\begin{equation}
K\,\mathbf{e}_\alpha \;=\; \lambda_\alpha\,\mathbf{e}_\alpha,
\end{equation}
where small $\lambda_\alpha$ correspond to large-scale, low-energy collective motions.  These modes manifest as out-of-phase displacements---one subset of arcs moves inward, another outward, preserving the knot.


\textbf{Dynamical Arrest at Large $n_c$.}---As the crossing number increases, more arcs share more crossing points, intensifying the off-diagonal coupling.  Eventually, the effective free-energy basins become \emph{so narrow} that even these collective modes are frozen. This distinction underscores the role of localized twisting in suppressing configurational entropy, ultimately leading to a more rigid and dynamically constrained polymer state. 
More generally, the \emph{loss of ergodicity} in complex knots establishes a connection between polymer topology and glassy dynamics, where frustration-driven constraints prevent full exploration of conformational space. 
This provides a new lens for understanding biological entanglement phenomena, such as the persistence of knotted DNA structures and the selective stability of protein loops with topological constraints.
Additionally, engineering applications of synthetic molecular knots could leverage these findings: by tuning crossing complexity, it may be possible to design materials with \emph{programmable dynamical arrest}, effectively controlling viscoelastic response through topology alone. Numerically, we observe the following:

\begin{enumerate}
\item \emph{Decreasing number of metastable states} $N_{\mathrm{states}}\sim A \exp\left(-\left(\frac{n_c}{\tau}\right)^\beta\right)$, reflecting how topological constraints suppress microstate accessibility. The introduction of the stretched exponential form captures the gradual reduction in microstate accessibility with increasing knot complexity $n_c$. Unlike a purely exponential decay, this model accounts for the initial sharp decline in metastable states at low complexity followed by a slower suppression at higher complexity, consistent with the emergence of cooperative modes and dynamical arrest observed in knotted ring polymers. Here, $A$ represents the initial number of states, $\tau$ defines the decay scale, and $\beta$ characterizes the stretching, indicating the nature of the cooperative effects driven by topological constraints.

This behavior closely parallels the dynamics of glass-forming materials, where stretched exponential decay (commonly described by the Kohlrausch-Williams-Watts function) characterizes the relaxation of systems as they approach a glassy state. In glasses, cooperative motions among particles become increasingly constrained due to growing dynamic heterogeneity, leading to a slowdown in relaxation processes. Similarly, in knotted ring polymers, topological constraints enforce cooperative arc motions, reducing the accessible conformations and driving the system toward a dynamically arrested state analogous to a glass transition.

\begin{figure}[h]
\centering
\includegraphics[width=1\linewidth]{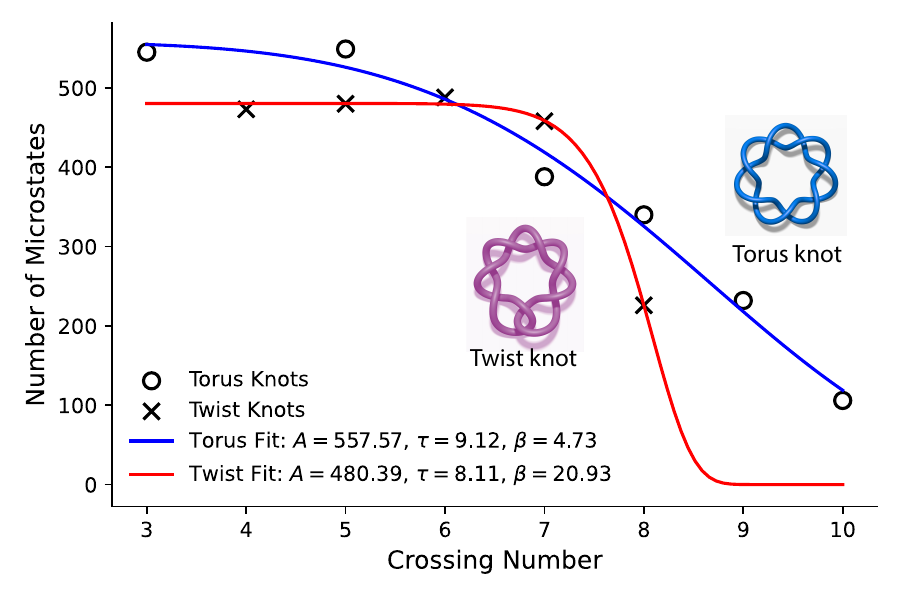}
\caption{Number of microstates from Molecular Dynamics trajectories as a function of the crossing number for torus knots (black circles) and twist knots (black crosses). The torus knot scaling trend (blue) and the twist knot trend (red) both follow stretched exponential fits, given by $N = A e^{-(x/\tau)^\beta}$. These fits capture the exponential suppression of microstates as crossing number increases, reflecting the increasing topological constraints on polymer configurational entropy. The difference in fit parameters highlights distinct relaxation behaviors: torus knots retain a higher number of microstates at a given crossing number, suggesting greater configurational flexibility, whereas twist knots experience a steeper decline, indicating stronger entropic confinement.}
\label{fig:stretched_exponential}
\end{figure}

Figure \ref{fig:stretched_exponential} illustrates the relationship between knot complexity and the number of metastable states, modeled using a stretched exponential function. The fit parameters ($A$, $\tau$, and $\beta$) provide a quantitative framework for understanding how topological constraints enforce dynamical arrest. The decrease in the number of microstates corresponds to the onset of cooperative arc motions, while the gradual flattening at higher complexity reflects the near-complete suppression of microstate accessibility. Torus knots exhibit more uniform topological constraints, while twist knots have localized constraints. As a result, the shape of the curve is different for twist knots, indicating an abrupt loss of configurational entropy. These fits quantify the suppression of microstates with increasing crossing number, reflecting the intensifying topological constraints on polymer configurational entropy. The difference in fit parameters highlights distinct relaxation behaviors between torus and twist knots. Torus knots (blue) retain a higher number of microstates at the same crossing number, indicating that their global structure permits greater configurational flexibility. This suggests that torus knots accommodate cooperative motion more effectively, with larger-scale conformational rearrangements remaining feasible.
Twist knots (red), by contrast, experience a steeper decline in microstates, demonstrating that their topological structure enforces stronger entropic confinement. The presence of localized twist-induced constraints disrupts cooperative motion, leading to faster topological arrest as crossing number increases. This distinction supports the central premise of the paper: topology dictates collective dynamics in knotted polymers. Torus knots, with their more delocalized crossings, exhibit weaker inter-arc couplings and a slower suppression of microstates. Twist knots, however, introduce localized constraints that induce cooperative displacement patterns at smaller length scales, reducing the number of accessible conformations more rapidly. These findings reinforce the arc-based model, which predicts emergent dynamical arrest at large crossing numbers as entropically favorable collective modes vanish.

\begin{figure}[h] \centering \includegraphics[width=1\linewidth]{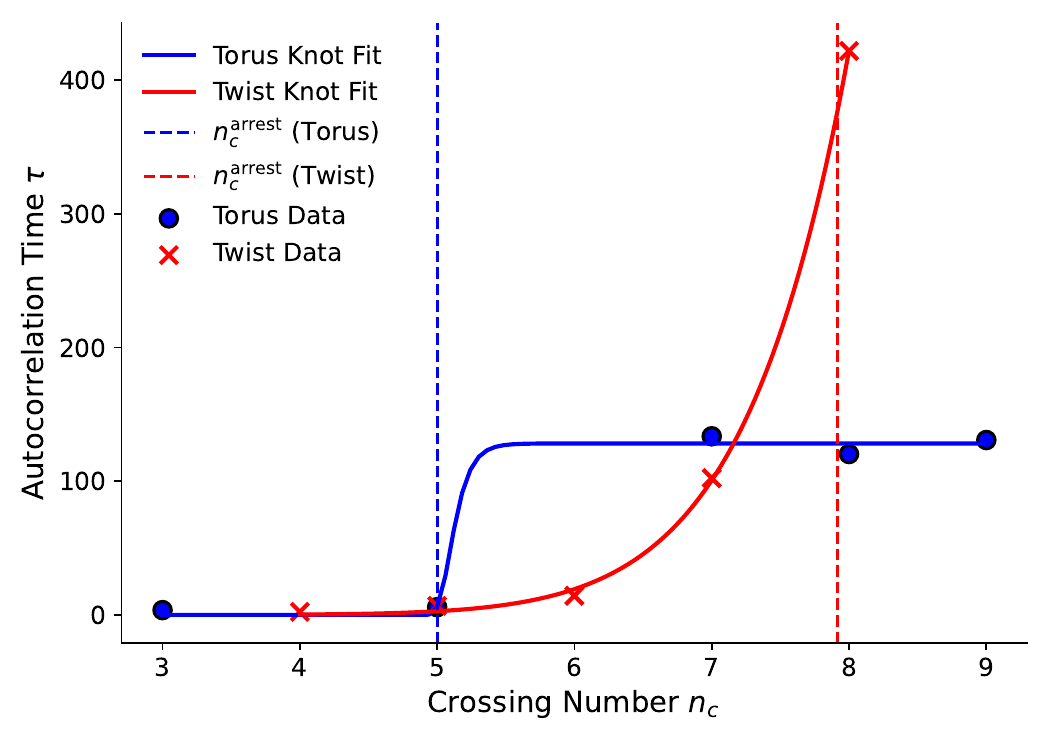} \caption{Autocorrelation time $\tau$ as a function of crossing number $n_c$ for torus (blue circles) and twist (red crosses) knots. The torus knots follow a stretched exponential scaling, while the twist knots obey a power-law dependence, highlighting distinct relaxation mechanisms. The dynamical arrest threshold $n_c^{\text{arrest}}$, where relaxation times increase sharply, is indicated by vertical dashed lines for each class. In torus knots, collective relaxation persists up to moderate complexity before topological constraints induce progressive arrest. In contrast, twist knots experience abrupt ergodicity breaking, where localized constraints restrict motion at significantly lower $n_c$. These differences illustrate two fundamentally distinct ergodicity-breaking pathways driven by polymer topology.}
 \label{fig:autocorrelation} \end{figure}
\item \emph{Collective motion} in moderate $n_c$ but suppressed at higher $n_c$; the knot is locked into \emph{dynamical arrest} with minimal fluctuations.
\end{enumerate}
Thus, \emph{only at moderate complexity} does one see significant cooperative modes.  Beyond that, large $n_c$ yields near-complete topological locking.


\textbf{Discussion.}---This analysis explains the \emph{qualitative} progression from low-complexity, loosely coordinated motion to high-complexity, strongly \emph{coupled} arcs and final topological locking.  Our \emph{novel contribution} is an explicit \emph{phenomenological} matrix formalism that treats each arc as a subchain and encodes crossing constraints in $K_{ij}$ or through Eq.\ \eqref{eq:constraint}.  Traditional Rouse- or Zimm-type models \cite{DoiEdwards, RubinsteinColby} do not include knotted topology or cooperative dynamics arising from topological constraints. Meanwhile, existing knot theories often focus on static properties such as geometry \cite{Katritch1996, Rawdon2015, Vologodskii1998} or ropelength arguments \cite{Grosberg2007} but do not address the dynamic behavior of \emph{cooperative arcs}. By incorporating dynamic coupling explicitly, our model bridges this gap, offering a unified framework for understanding topological constraints and their effects on motion. Our results also reveal a topology-driven divergence in autocorrelation scaling laws between torus and twist knots. The stretched exponential dependence in torus knots suggests that entropic constraints emerge gradually, allowing correlated collective motion at moderate crossing numbers. In contrast, the power-law scaling in twist knots indicates that dynamical arrest occurs more abruptly, driven by local constraints that disrupt larger-scale collective relaxations. This fundamental distinction is quantitatively confirmed in Figure \ref{fig:autocorrelation}, which illustrates the topology-dependent divergence in autocorrelation time. Torus knots exhibit a gradual, stretched-exponential increase in relaxation time, while twist knots display a much steeper power-law growth, indicating faster configurational trapping. This distinction is quantitatively confirmed in Figure \ref{fig:autocorrelation}, which illustrates the divergence in autocorrelation time scaling. Torus knots exhibit a stretched-exponential growth, while twist knots display a steep power-law increase, signaling faster configurational trapping.

We anticipate applications for knotted DNA, synthetic molecular knots, and protein loops \cite{Yeates2012, Orlandini2016, Zhang2019}, where experimental observations of ``partial arrest'' or slow correlated domain motion could be rationalized by strong off-diagonal arc couplings.  Furthermore, the stretched exponential relationship between the number of metastable states and knot complexity provides a quantitative basis for interpreting dynamical arrest in these systems. Future refinements might include fully \emph{nonlinear} forms of $F_{\mathrm{coll}}$ to capture large deformations, or time-dependent topological transitions under external forcing, analogous to relaxation phenomena in glasses.


\textbf{Conclusions.}---We have introduced a new, \emph{arc-based} phenomenological theory to explain the emergence of \emph{collective} (coordinated) modes in knotted ring polymers and their subsequent \emph{dynamical arrest} at high crossing number.  By explicitly linking each arc's displacement to the knot's crossing constraints, we obtain a coupling matrix whose off-diagonal elements grow with knot complexity.  This yields cooperative motion at moderate complexity and near-complete locking at large $n_c$, consistent with the dynamical arrest trends observed in Figure\ref{fig:autocorrelation}.
  Our approach paves the way for quantitative modeling of topologically driven collective dynamics in a broad class of \emph{complex} soft matter systems.


\begin{acknowledgments}
The author thanks Eugene Terentjev for inspiring discussions and acknowledges the generous support from the U.S. Department of Energy, Office of Science, through the Biological and Environmental Research (BER) and the Advanced Scientific Computing Research (ASCR) programs under contract number 89233218CNA000001 and NSF (MCB 2337393).
\end{acknowledgments}


\end{document}